\input harvmac\skip0=\baselineskip

\def\encadremath#1{\vbox{\hrule\hbox{\vrule\kern8pt\vbox{\kern8pt
\hbox{$\displaystyle #1$}\kern8pt} \kern8pt\vrule}\hrule}}

\def\p{\partial}
\def\rl{{SL(2,{\bf R})}}
\def\su{SU(2)}
\def\osp{Osp(4^*|4)}



\lref\AntoniadisEG{
  I.~Antoniadis, S.~Ferrara, R.~Minasian and K.~S.~Narain,
  ``$R^4$ couplings in M- and type II theories on Calabi-Yau spaces,''
  Nucl.\ Phys.\  B {\bf 507}, 571 (1997)
  [arXiv:hep-th/9707013].
}

\lref\FrappatPB{
  L.~Frappat, P.~Sorba and A.~Sciarrino,
  ``Dictionary on Lie superalgebras,''
  arXiv:hep-th/9607161.
}
\lref\AwadaEP{
  M.~Awada and P.~K.~Townsend,
  ``N=4 Maxwell-Einstein Supergravity In Five-Dimensions And Its SU(2)
  Gauging,''
  Nucl.\ Phys.\  B {\bf 255}, 617 (1985).
}

\lref\BerensteinGJ{
  D.~Berenstein and R.~G.~Leigh,
  ``Spacetime supersymmetry in AdS(3) backgrounds,''
  Phys.\ Lett.\  B {\bf 458}, 297 (1999)
  [arXiv:hep-th/9904040].
}

\lref\BergshoeffHC{
  E.~Bergshoeff, S.~Cucu, M.~Derix, T.~de Wit, R.~Halbersma and A.~Van Proeyen,
  ``Weyl multiplets of N = 2 conformal supergravity in five dimensions,''
  JHEP {\bf 0106}, 051 (2001)
  [arXiv:hep-th/0104113].
}

\lref\BergshoeffKH{
  E.~Bergshoeff, S.~Cucu, T.~de Wit, J.~Gheerardyn, S.~Vandoren and A.~Van Proeyen,
  ``N = 2 supergravity in five dimensions revisited,''
  Class.\ Quant.\ Grav.\  {\bf 21}, 3015 (2004)
  [Class.\ Quant.\ Grav.\  {\bf 23}, 7149 (2006)]
  [arXiv:hep-th/0403045].
}

\lref\BershadskyMS{
  M.~A.~Bershadsky,
  ``Superconformal algebras in two dimensions with arbitrary N,''
  Phys.\ Lett.\  B {\bf 174}, 285 (1986).
}

\lref\BowcockBM{
  P.~Bowcock,
  ``Exceptional Superconformal Algebras,''
  Nucl.\ Phys.\  B {\bf 381}, 415 (1992)
  [arXiv:hep-th/9202061].
}

\lref\BrittoPacumioAX{
  R.~Britto-Pacumio, J.~Michelson, A.~Strominger and A.~Volovich,
  ``Lectures on superconformal quantum mechanics and multi-black hole  moduli
  arXiv:hep-th/9911066.
}

\lref\CallanDJ{
  C.~G.~.~Callan, J.~A.~Harvey and A.~Strominger,
  ``World sheet approach to heterotic instantons and solitons,''
  Nucl.\ Phys.\  B {\bf 359}, 611 (1991).
}

\lref\CastroHC{
  A.~Castro, J.~L.~Davis, P.~Kraus and F.~Larsen,
  ``5D Black Holes and Strings with Higher Derivatives,''
  arXiv:hep-th/0703087.
}

\lref\CastroSD{
  A.~Castro, J.~L.~Davis, P.~Kraus and F.~Larsen,
  ``5D attractors with higher derivatives,''
  JHEP {\bf 0704}, 091 (2007)
  [arXiv:hep-th/0702072].
}

\lref\CastroCI{
  A.~Castro, J.~L.~Davis, P.~Kraus and F.~Larsen,
  ``Precision entropy of spinning black holes,''
  arXiv:0705.1847 [hep-th].
}

\lref\asdh{A. Dabholkar and A. Strominger, unpublished (2005).}
\lref\DabholkarGP{
  A.~Dabholkar and S.~Murthy,
  ``Fundamental Superstrings as Holograms,''
  arXiv:hep-th/07073818.
}

\lref\DabholkarDQ{
  A.~Dabholkar, R.~Kallosh and A.~Maloney,
  ``A stringy cloak for a classical singularity,''
  JHEP {\bf 0412}, 059 (2004)
  [arXiv:hep-th/0410076].
}

\lref\DabholkarYR{
  A.~Dabholkar,
  ``Exact counting of black hole microstates,''
  Phys.\ Rev.\ Lett.\  {\bf 94}, 241301 (2005)
  [arXiv:hep-th/0409148].
}

\lref\DabholkarNC{
  A.~Dabholkar, J.~P.~Gauntlett, J.~A.~Harvey and D.~Waldram,
  ``Strings as Solitons and Black Holes as Strings,''
  Nucl.\ Phys.\  B {\bf 474}, 85 (1996)
  [arXiv:hep-th/9511053].
}

\lref\DabholkarYF{
  A.~Dabholkar, G.~W.~Gibbons, J.~A.~Harvey and F.~Ruiz Ruiz,
  ``Superstrings and solitons,''
  Nucl.\ Phys.\  B {\bf 340}, 33 (1990).
}

\lref\deWitPX{
  B.~de Wit, P.~G.~Lauwers and A.~Van Proeyen,
  ``Lagrangians Of N=2 Supergravity - Matter Systems,''
  Nucl.\ Phys.\  B {\bf 255}, 569 (1985).
}

\lref\FerraraHH{
  S.~Ferrara, R.~R.~Khuri and R.~Minasian,
  ``M-Theory on a Calabi-Yau Manifold,''
  Phys.\ Lett.\  B {\bf 375}, 81 (1996)
  [arXiv:hep-th/9602102].
}

\lref\FradkinBZ{
  E.~S.~Fradkin and V.~Y.~Linetsky,
  ``Results Of The Classification Of Superconformal Algebras In
  Two-Dimensions,''
  Phys.\ Lett.\  B {\bf 282}, 352 (1992)
  [arXiv:hep-th/9203045].
}

\lref\FradkinCB{
  E.~S.~Fradkin and V.~Y.~Linetsky,
  ``Classification Of Superconformal Algebras With Quadratic Nonlinearity,''
  arXiv:hep-th/9207035.
}

\lref\FradkinGJ{
  E.~S.~Fradkin and V.~Y.~Linetsky,
  ``An Exceptional N=8 superconformal algebra in two-dimensions associated with
  F(4),''
  Phys.\ Lett.\  B {\bf 275}, 345 (1992).
}

\lref\FradkinKM{
  E.~S.~Fradkin and V.~Y.~Linetsky,
  ``Classification Of Superconformal And Quasisuperconformal Algebras In
  Two-Dimensions,''
  Phys.\ Lett.\  B {\bf 291}, 71 (1992).
}

\lref\FujitaKV{
  T.~Fujita and K.~Ohashi,
  ``Superconformal tensor calculus in five dimensions,''
  Prog.\ Theor.\ Phys.\  {\bf 106}, 221 (2001)
  [arXiv:hep-th/0104130].
}

\lref\GarfinkleQJ{
  D.~Garfinkle, G.~T.~Horowitz and A.~Strominger,
  ``Charged black holes in string theory,''
  Phys.\ Rev.\  D {\bf 43}, 3140 (1991)
  [Erratum-ibid.\  D {\bf 45}, 3888 (1992)].
}

\lref\GiddingsWN{
  S.~B.~Giddings, J.~Polchinski and A.~Strominger,
  ``Four-dimensional black holes in string theory,''
  Phys.\ Rev.\  D {\bf 48}, 5784 (1993)
  [arXiv:hep-th/9305083].
}

\lref\GiveonJG{
  A.~Giveon and M.~Rocek,
  ``Supersymmetric string vacua on AdS(3) x N,''
  JHEP {\bf 9904}, 019 (1999)
  [arXiv:hep-th/9904024].
}

\lref\GiveonKU{
  A.~Giveon and A.~Pakman,
  ``More on superstrings in AdS(3) x N,''
  JHEP {\bf 0303}, 056 (2003)
  [arXiv:hep-th/0302217].
}

\lref\GiveonNS{
  A.~Giveon, D.~Kutasov and N.~Seiberg,
  ``Comments on string theory on AdS(3),''
  Adv.\ Theor.\ Math.\ Phys.\  {\bf 2}, 733 (1998)
  [arXiv:hep-th/9806194].
}

\lref\GiveonPR{
  A.~Giveon and D.~Kutasov,
  ``Fundamental strings and black holes,''
  JHEP {\bf 0701}, 071 (2007)
  [arXiv:hep-th/0611062].
}
\lref\DabholkarDT{
  A.~Dabholkar, F.~Denef, G.~W.~Moore and B.~Pioline,
  ``Precision counting of small black holes,''
  JHEP {\bf 0510}, 096 (2005)
  [arXiv:hep-th/0507014].
}
\lref\DabholkarBY{
  A.~Dabholkar, F.~Denef, G.~W.~Moore and B.~Pioline,
  ``Exact and asymptotic degeneracies of small black holes,''
  JHEP {\bf 0508}, 021 (2005)
  [arXiv:hep-th/0502157].
}
\lref\HanakiPJ{
  K.~Hanaki, K.~Ohashi and Y.~Tachikawa,
  ``Supersymmetric Completion of an $R^2$ Term in Five-Dimensional
  Supergravity,''
  Prog.\ Theor.\ Phys.\  {\bf 117}, 533 (2007)
  [arXiv:hep-th/0611329].
}
\lref\clifford{ Clifford V. Johnson , " Heterotic Coset Models of Microscopic Strings and Black Holes", arXiv:hep-th/07074303.}

\lref\HenneauxIB{
  M.~Henneaux, L.~Maoz and A.~Schwimmer,
  ``Asymptotic dynamics and asymptotic symmetries of three-dimensional
  extended AdS supergravity,''
  Annals Phys.\  {\bf 282}, 31 (2000)
  [arXiv:hep-th/9910013].
}

\lref\HorowitzCD{
  G.~T.~Horowitz and A.~Strominger,
  ``Black strings and P-branes,''
  Nucl.\ Phys.\  B {\bf 360}, 197 (1991).
}
\lref\StromingerKF{
  A.~Strominger,
  ``Macroscopic Entropy of $N=2$ Extremal Black Holes,''
  Phys.\ Lett.\  B {\bf 383}, 39 (1996)
  [arXiv:hep-th/9602111].
}
\lref\ItoVD{
  K.~Ito,
  ``Extended superconformal algebras on AdS(3),''
  Phys.\ Lett.\  B {\bf 449}, 48 (1999)
  [arXiv:hep-th/9811002].
}

\lref\KnizhnikWC{
  V.~G.~Knizhnik,
  ``Superconformal algebras in two dimensions,''
  Theor.\ Math.\ Phys.\  {\bf 66}, 68 (1986)
  [Teor.\ Mat.\ Fiz.\  {\bf 66}, 102 (1986)].
}

\lref\KutasovXU{
  D.~Kutasov and N.~Seiberg,
  ``More comments on string theory on AdS(3),''
  JHEP {\bf 9904}, 008 (1999)
  [arXiv:hep-th/9903219].
}

\lref\KutasovZH{
  D.~Kutasov, F.~Larsen and R.~G.~Leigh,
  ``String theory in magnetic monopole backgrounds,''
  Nucl.\ Phys.\  B {\bf 550}, 183 (1999)
  [arXiv:hep-th/9812027].
}

\lref\SenKJ{
  A.~Sen,
  ``Stretching the horizon of a higher dimensional small black hole,''
  JHEP {\bf 0507}, 073 (2005)
  [arXiv:hep-th/0505122].
}

\lref\BrittoPacumioAX{
  R.~Britto-Pacumio, J.~Michelson, A.~Strominger and A.~Volovich,
  ``Lectures on superconformal quantum mechanics and multi-black hole  moduli
  spaces,''
  arXiv:hep-th/9911066.
}

\lref\BrownNW{
  J.~D.~Brown and M.~Henneaux,
  ``Central Charges in the Canonical Realization of Asymptotic Symmetries: An
  Example from Three-Dimensional Gravity,''
  Commun.\ Math.\ Phys.\  {\bf 104}, 207 (1986).
}

\lref\ZamolodchikovWN{
  A.~B.~Zamolodchikov,
  ``Infinite Additional Symmetries In Two-Dimensional Conformal Quantum Field
  Theory,''
  Theor.\ Math.\ Phys.\  {\bf 65}, 1205 (1985)
  [Teor.\ Mat.\ Fiz.\  {\bf 65}, 347 (1985)].
}

\lref\as{J. M. Lapan, A. Simons and A.  Strominger, ``Search for the dual of N heterotic strings'', http://gesalerico.ft.uam.es/strings07, presentation by A. Strominger at  Strings '07, Madrid,  June, 2007.}
\lref\ChamseddinePI{
  A.~H.~Chamseddine, S.~Ferrara, G.~W.~Gibbons and R.~Kallosh,
  ``Enhancement of supersymmetry near 5d black hole horizon,''
  Phys.\ Rev.\  D {\bf 55}, 3647 (1997)
  [arXiv:hep-th/9610155].
}

\def\apm{{\alpha^{\prime}}}


\Title{\vbox{\baselineskip12pt\hbox{} }} { Nearing the Horizon  of a
Heterotic String} \centerline{Joshua M. Lapan, Aaron Simons and Andrew
Strominger} 
\bigskip
\centerline{Center for the Fundamental Laws of Nature}\centerline{ Jefferson Physical
Laboratory, Harvard University, Cambridge, MA  02138}
\smallskip
\centerline{} \vskip .6in \centerline{\bf Abstract} {It is argued that recent developments point to the existence of an $AdS_3\times S^2\times T^5$ holographic dual for the 2D CFT living on the worldsheet of $N$ coincident heterotic strings in a $T^5$ compactification, which can in turn be described by an exact worldsheet CFT.  A supergravity analysis is shown to imply that the global supergroup is $\osp$, with 16 supercharges and an  affine extension given, surprisingly,  by a {\it nonlinear } ${\cal N}=8$ 2D superconformal algebra.  Possible supergroups with 16 supercharges are also found to match the expected symmetries for $T^n$ compactification with $0\le n\le 7$. 
} \vskip .3in

\Date{} \listtoc \writetoc

\newsec{Introduction and summary}
The worldsheet of $N$ stretched, coincident heterotic strings is
described by a $(c_L,c_R)=(24N,12N)$ 2d CFT. General considerations as well as recent investigations, both described below,
raise the intriguing possibility that this CFT has an
$AdS_3\times M $ holographic spacetime dual. If so, the
first-quantized Hilbert space of the $N$ stretched heterotic strings
would be identified with the second-quantized Hilbert space of
interacting closed heterotic strings on $AdS_3\times M$.  In this paper, as reported in \as, we continue these investigations and in particular find some surprising results about the structure of the supersymmetry group. 

  Why should we expect such a holographic dual? At strong coupling, the
heterotic string becomes a $D1$-brane of the type I theory. General
low-energy scaling arguments coupled with open-closed duality then suggest the
existence of a holographic dual. Because the low-energy limit of the
worldvolume theory is conformally invariant, the dual should contain
an $AdS_3$ factor. In addition, $N$ stretched heterotic strings have
an exponentially growing spectrum of left-moving BPS excitations.
Although the growth is not enough to make a black string with a
horizon large compared to the string scale, it is still the case
that in the classical limit,\foot{With the spacetime momentum
density along the string, and all spacetime fields fixed while $\hbar
 \to 0$.} the second law of thermodynamics forbids energy from
leaking off of the $N$ strings, just as it does for a large black hole. One expects this behavior to be
explained in the macroscopic spacetime picture by the appearance of
a stringy horizon and associated near-horizon scaling solution.
However, the real situation is likely more subtle than these general comments indicate.  As we shall see below,
simple group theory implies that the situation is highly
dimension-dependent. In particular, concrete indicators of a
holographic dual in the special case of compactification to $D=5$, on which we largely concentrate, 
will be reviewed below. 

\subsec{The leading-order solution} The string frame 
classical geometry sourced by the $N$ stretched heterotic strings in the leading $\apm$ approximation for
a compactification to $D\ge 5$ dimensions was found some time ago
\refs{\DabholkarYF} using the supergravity equations:
\eqn\njk{ds^2={dx^+dx^- \over 1+{N ({r_h\over r})^{D-4}}}-d \vec x \cdot d
\vec x-ds_{10-D}^2,}  where $r_h^{D-4}={g_{10}^2 \over 8 \pi^5 V_{10-D}}$, $\vec x$ is a transverse $D-2$ vector
and $r^2= \vec x \cdot \vec x$. The string coupling behaves as
\eqn\jky{e^{2\Phi}={e^{2\Phi_0}\over 1+N ({r_h \over r})^{D-4}},} and there is
also a Kalb-Ramond  field \eqn\fjh{H=dx^+dx^-de^{2(\Phi-\Phi_0)}.} This
spacetime is singular at the core of the string $r=0$. Interestingly
the string coupling goes to zero while the curvature diverges. This
suggests the possibility that the singularity might be resolved
within classical string theory by $\apm$ corrections. \subsec{Small
4d black holes and small 5d black strings} Recently there have been
compelling indicators \refs{\DabholkarYR\DabholkarDT\DabholkarBY \SenKJ  \DabholkarDQ  \CastroHC\CastroSD-\CastroCI} that such a stringy resolution
in fact occurs for the case $D=5$. The story began with an $S^1$
compactification from $5$ to $4$ dimensions, in which the stretched
strings are wrapped around the $S^1$ and become a pointlike object
in $D=4$. This ``small black hole" has BPS excitations with
momentum-winding $(N,k)$ and a degeneracy that grows at large
charges as $e^{4\pi \sqrt{Nk}}$. As above, this growth is not rapid
enough to make a large black hole visible in supergravity, for which
it is easily seen that the entropy must scale as the square of the
charges. Nevertheless, the charges of the small black hole were
plugged into the entropy formula derived in an $\apm$ expansion for
large black holes and found to reproduce -- to all orders! -- the known
BPS degeneracies \refs{\DabholkarYR\DabholkarDT-\DabholkarBY}.

This impressive agreement was surprising because the macroscopic
derivation of the entropy as a function of the charges employs the
known spacetime black hole attractor geometry \StromingerKF\ as an
intermediate step. For small black holes, no such solutions were
known. Subsequently, it was found that when stringy $R^2$ corrections
to the supergravity equations are included, solutions with
string-scale horizon do exist, and furthermore the horizon area
scales in the right way with the charges \SenKJ. Of course such
solutions can only be regarded as suggestive, due to the ambiguities
arising from field redefinitions and uncontrolled effects of $R^4$ and
higher corrections. Nevertheless, the remarkable coherence of the
small black hole picture suggests that we take them seriously. Ultimately
the existence or not of these solutions should be addressed using
worldsheet CFT methods, which can control all the $\apm$
corrections.

\subsec{Small black strings} The near-horizon geometry
of the small black holes contains an $AdS_2$ factor with an electric
field associated to the Kaluza-Klein $U(1)$. Hence we have an $S^1$
fibered over the $AdS_2$. The total space of such a bundle is a
quotient of $AdS_3$. Taking the cover of this quotient, we obtain
the $AdS_3$ factor of the near horizon geometry of $N$ stretched
heterotic strings in $5$ dimensions. Indeed the full 5d solutions
were seen directly in $T^5$ compactification to $5$ dimensions in a
recent elegant paper CDKL (Castro, Davis, Kraus and Larsen) \CastroHC. CDKL begin with the $R^2$-corrected $D=5$ ${\cal N}=2$
supersymmetry transformation laws and find half BPS solutions with a near-horizon
$AdS_3\times S^2$ factor and charges corresponding to $N$ heterotic
strings. In the heterotic frame, these solutions have a string
coupling proportional to ${ 1 \over \sqrt{N}}$.

\subsec{The near-horizon nonlinear superconformal group} The
symmetries of the near-horizon region of this solution are of special
interest. One expects the number of supersymmetries in the near
horizon region to double from 8 to 16. But there are only four Lie
superalgebras of classical type with 16 supercharges and an $SL(2,{\bf R})$ factor: $\osp,~~SU(1,1|4),~~F(4)$, and $Osp(8|2)$,
with bosonic R-symmetry factors $SU(2)\times Sp(4),~~SU(4)\times
U(1),~~ SO(7)$, and $SO(8)$, respectively.\foot{ See for example \FrappatPB. One may also consider the product group $D(2,1;\alpha)\times D(2,1;\alpha)$. } This is puzzling for two
reasons. Firstly, R-symmetries usually arise geometrically as
spacetime isometries -- {\it e.g.} the $SO(6)$ R-symmetry corresponding to
$S^5$ rotations of the near horizon D3 geometry. But in the current
context that gives at most the $SU(2)$ rotations of the $S^2$ and so cannot account for the R-symmetry of any of the above supergroups.
Secondly, as shown by Brown and Henneaux \BrownNW, when there is an
$AdS_3$ factor the superisometry group must have an affine extension
containing a Virasoro algebra. However there are no linear ``${\cal N}=8$"
superconformal algebras containing any of the above superalgebras as
global subalgebras.

We show herein by direct computation that the global superisometry group is
in fact $\osp$. As usual, the $SU(2)$ factor of the R-symmetry
arises from the geometric rotational isometries of the $S^2$
R-symmetry. From the 5d point of view, the $Sp(4)\sim SO(5)$ arises
unusually from the global $Sp(4)$ R-symmetry of 5d ${\cal N}=4$
supergravity. From the 10d point of view, these are the $SO(5)$
rotations of the spin frame for the $T^5$. This $SO(5)$ acts only on
fermions and is not to be confused with spacetime rotations.

While this explains how the large global R-symmetry arises, it does not explain the puzzle with
the affine extension. While there are no linear
superconformal algebras with more than 4 supercurrents (which means
8 global supercharges in the NS sector), there are a few {\it nonlinear}
algebras with 8 supercurrents. These were classified some time ago
by \refs{\KnizhnikWC ,\BershadskyMS ,\BowcockBM}, and one of these algebras, let us denote it
$\hat \osp$, indeed contains $\osp$ in the $k \to \infty$ limit. The nonlinearity in the
commutation relation is confined to the commutator of the
supercurrents, which takes the schematic form \eqn\rfvb{ \{
G^I_r,G^J_s\}\sim 2\delta^{IJ}L_{r+s}+(r-s)(R^{IJ})_{r+s}+\sum_{p}
(R^{I}{}_K)_{r+s-p}(R^{KJ})_p+\cdots,} where the
current $R^{IJ}$ generates the bosonic R-symmetry group.
The nonlinear superconformal algebras are a special type of $W$
algebra with only one spin two current and no higher currents.
Though known for some time \ZamolodchikovWN,
these algebras have not seen many
applications in string theory or elsewhere. We note that it is not
fully understood when these algebras have unitary
representations.

Fortuitously, consistent boundary conditions on $AdS_3$ with more
than eight global supersymmetries and their associated asymptotic
symmetry algebras were studied in \HenneauxIB. The short list contains
$\hat \osp$. We conclude that the near-horizon symmetry algebra of the
$R^2$-corrected supergravity solutions corresponding to $N$ stretched
heterotic strings is $\hat \osp$. \subsec{ $D\neq 5$} It is
interesting to see how or if a picture could emerge in dimensions
other than 5 consistent with the the known supergroups. Near-horizon
symmetry enhancement suggests that there should always be 16
near-horizon supersymmetries.\foot{14 is another possibility,
corresponding to the supergroups $G(3)$ with R-symmetry group $G_2$ or $Osp(7|2)$ with $SO(7)$.}
In $D=10$, it is natural to speculate that there is a
stretched-string solution with an $AdS_3\times S^7$ near horizon
region with the $Osp(8|2)$ superisometry group and a geometrically
realized $SO(8)$ R-symmetry. In $D=9$, $F(4)$ could arise with a
geometrical $SO(7)$. It could also arise in $D=3$, but with a
nongeometrical $SO(7)$ from spin frame rotations of the $T^7$. In
$D=8$, the horizon is an $S^5$, so we could have $SU(1,1|4)$ with
the $SU(4)\sim SO(6)$ geometrically realized and the $U(1)$
nongeometrical. $SU(1,1|4)$ is also a candidate for $D=4$ with a
nongeometrical $SO(6)$ and the $U(1)$ realized geometrically as
rotations of the $S^1$ horizon. In $D=7$, the horizon is an $S^4$ so
one could again have $\osp$, but with a geometrical
$Sp(4)\sim SO(5)$ and a nongeometrical $SU(2)$. In $D=6$, which is
the self-dual dimension for strings, the near horizon geometry would
be $AdS_3\times S^3\times T^4$. This has both a geometrical and a
nongeometrical $SO(4)$, both of which have $SU(2)$ subgroups. This
could correspond to two copies of $D(2,1;\alpha)$ with left and right actions, each containing an $SU(2)\times SU(2) $
R-symmetry. So for all $3\le D \le 10$, there are candidate
near-horizon supergroups with 16 supercharges.\foot{Candidates for near-horizon supergroups of type II strings can be obtained by taking left and right copies of the above, except for the case $D=6$.} Whether or not the
solutions actually exist remains to be seen.

\subsec{A worldsheet CFT?}

 As discussed above, the success of the small black hole/string story
 suggested that $N$ heterotic strings in $D=5$ have an $AdS_3\times
 S^2$ near horizon region. The value of the string coupling goes to
 zero as $N\to \infty$ so that string loop corrections can be ignored.
 Such solutions were then found in the classical stringy $R^2$-corrected supergravity, but $R^2$-corrected supergravity
 is unreliable because $\apm$ corrections are uncontrolled. Such
 corrections, however, are controllable using worldsheet CFT methods, so either the authors of \refs{\SenKJ, \DabholkarDQ, \CastroHC} as well  as the authors of the present
 paper were misled by the solutions of $R^2$-corrected supergravity,
 or an exact worldsheet CFT which describes the near-horizon
 geometry must exist.

 The sought after worldsheet CFT cannot involve a RR background, as
 there are none in heterotic string theory. Furthermore, the large
 spacetime symmetry group places strong constraints on the
 worldsheet CFT \refs{\GiveonNS, \KutasovXU,\GiveonJG, \GiveonKU}. So if this CFT and the associated GSO projection exist, it should be possible
 to find them. Related and in some cases partial proposals have already appeared in \refs{\asdh, \GiveonPR, \KutasovZH} as well as \refs{\DabholkarGP, \clifford} which appeared as the present work was under submission. 
 The closely related problem of finding the CFT which describes the $S^2$ horizon of a heterotic monopole was solved in \GiddingsWN. We will review this construction and its application to the current problem in the last section. 

 Supposing the CFT does not exist for some or all cases, and we have simply been misled
 by the $R^2$ solutions, what are the possibilities?
 One is that there simply is no near horizon solution and that both
 supergravity and the exact classical string theory are singular at
 the core of the string. A second possibility, advocated in \GiveonPR\ (but at odds with the picture in CDKL), is that there is a
 smooth near horizon solution, but that some of the supersymmetries
 act trivially. Phenomena of this type
 are known in string theory. For example if we look at the
 magnetically charged black hole solutions of \GiddingsWN, for magnetic
 charge $\pm 1$ the horizon part of the ``throat" theory is trivial
 and $SO(3)$ spacetime rotations act trivially. Should this turn out
 to be the case we will need to understand in what sense the
 near-horizon spacetime is the holographic dual of the heterotic
 string CFT.

\newsec{Near horizon analysis}

In this section we explicitly demonstrate, by finding the unbroken supersymmetries and computing their commutators,  that  the superisometry group of  the near-horizon region of a fundamental heterotic string in $R^2$-corrected supergravity is $\osp$. We employ the asymptotically flat solution of the BPS conditions found in  CDKL.  The CDKL analysis was in turn made possible by the recent supersymmetric completion of the relevant $R^2$ term in five dimensions \HanakiPJ , which descends from terms related to  anomaly cancellation in the M-theory lift. 
\subsec{Supergravity in 5d}

Five dimensional supergravity with $8n$ real supercharges, conventionally referred to as ${\cal N}=2n$ supergravity, has an  $Sp(2n)$ R-symmetry group with the supersymmetry parameter $\epsilon^i,~~i=1,\dots ,2n,$ transforming in the  ${\bf { 2n}}$.  CDKL work in an offshell ${\cal N}=2$ formalism (which greatly simplifies the computation), but their solution can be embedded in an ${\cal N}=4$ theory as follows. The ${\cal N}=4$ gravitino variation has relevant terms of the form  \eqn\gengv{\delta \psi^i_\mu \sim{\nabla}_\mu \epsilon^i+ 
( F_{\rho\sigma}^{ij}+
G_{\rho\sigma} \Omega^{ij} ) ( \gamma_\mu {}^{\rho\sigma} - 4\delta_\mu {}^\rho \gamma^\sigma ) \epsilon_j+\cdots ,} where $F^{ij}$ and $G$ are 2-form field strengths in the ${\bf 5}$ and ${\bf 1}$ of $Sp(4)$ respectively. However, one can see upon dimensional reduction from $D=10$ that the $F^{ij}$ come from components of the metric $g$ and anti-symmetric 2-form $B$ which are mixed between $AdS_3 \times S^2$ and $T^5$, and these vanish in the present context. Under these circumstances, the $Sp(4)$ R symmetry is unbroken by the background and the ${\cal N}=4$ variation looks exactly like that of ${\cal N}=2$ but with $i=1,\dots,4$, instead of $i=1,2$. 

In $4+1$ dimensions there is no ordinary Majorana condition, but one can impose a symplectic-Majorana condition via \eqn\symmaj{{\bar \xi}^i = \xi_i^\dagger \gamma^{\hat 0}=\xi^{iT}C,} where $C$ is the charge conjugation matrix and tangent-space indices are hatted. We will also use the symplectic matrix $\Omega_{ij}$ to raise and lower indices by \eqn\nwse{\xi^i=\Omega^{ij}\xi_j \qquad \xi_i=\xi^j\Omega_{ji} .} We choose a basis in which \eqn\omga{\Omega_{12}=\Omega_{34}=-\Omega_{21}=-\Omega_{43}=1.} The string solution in supergravity has $ISO(1,1)\times SO(3)$ isometry. It is convenient to choose lightcone coordinates along the string $x^{\pm}=x^0 \pm x^1$, and spherical coordinates $r$,$\theta$,$\phi$ for the transverse directions. In conformity with CDKL, we take the tangent space metric to have signature $(+----)$.

We work in a representation of the Clifford algebra with $\gamma^{\hat 0}$ Hermitian, and the other $\gamma^{\hat \mu}$ anti-Hermitian. Consistent with this, we can choose $\gamma^{\hat 1}$ to be real while the others are pure imaginary. The charge conjugation matrix $C=\gamma^{{\hat 0}{\hat 1}}$ satisfies \eqn\cid{C\gamma^\mu C^{-1}=\gamma^{\mu T}.} As this is a non-chiral theory, we choose $\gamma^{\hat 0\hat 1\hat r\hat\theta\hat\phi}=1$ where the indices are tangent space indices, raised and lowered by $-\eta_{\hat 0\hat 0}=\eta_{\hat 1\hat 1}=\eta_{\hat r\hat r}=\eta_{\hat \theta\hat  \theta}=\eta_{\hat \phi\hat \phi}=-1$. Note that $\gamma^{\mu_1 \dots \mu_p}$ is always the antisymmetric combination divided by $p!$.

\subsec{Killing spinors}
The CDKL solution has an $AdS_3 \times S^2$ near horizon region with metric \eqn\nhmet{ ds^2={r \over l} dx^+ dx^- - {l^2 \over r^2} dr^2-l^2 d\Omega_2.} 
Choosing the vielbein \eqn\vbein{e^{\hat +}{}_{+}=e^{\hat -}{}_{-}=\sqrt{r \over l}, \qquad e^{\hat r}{}_{r}={l\over r}, \qquad e^{\hat \theta}{}_{\theta}=l, \qquad e^{\hat \phi}{}_{\phi}=l \sin \theta,} the only non-zero components of the spin connection are \eqn\spinc{\omega_{\phi}{}^{\hat \theta\hat \phi}=\cos \theta, \qquad \omega_{+}{}^{\hat r\hat +}=\omega_{-}{}^{\hat r\hat -}={1\over 2}\sqrt{r \over l^3}.} The Weyl multiplet of $5d$ ${\cal N}=2$ conformal supergravity contains an auxiliary $2$-form $v_{\mu\nu}$ (related to $G$ in \gengv) which is $v_{\hat \theta\hat \phi}={3\over 4l}$ in this background.  In terms of $v$ the precise version of the gravitino variation \gengv\ is  \eqn\grav{\delta_\epsilon \psi^i_{\mu}=( \nabla_{\mu} + {1\over 2}v_{\nu\rho}\gamma^{\nu\rho}{}_{\mu}-{1\over 3}v_{\nu\rho}\gamma_{\mu} \gamma^{\nu\rho} )\epsilon^i.} As mentioned above, because our background preserves R-symmery the R-symmetry index just goes along for the ride. There is also a second fermion $\chi$ as well as gauginos whose variations determine the scalar auxilliary field $D$ and field strengths, but turn out to not  further constrain the Killing spinor and so shall not concern us here. 

Let's first consider the $\delta \psi^i_r$ variation. There are two solutions with $r$-dependence $(r/l)^{\pm 1/4}$ and satisfyng the projection $\gamma^{\hat r\hat \theta\hat \phi}\epsilon^i_\pm= \pm \epsilon_\pm^i$.    Further, the two solutions are related by 
$\epsilon^i_-=(\sqrt{l/r})\gamma^{\hat +\hat r}\epsilon_+^i$.  
Denote  by $\epsilon^i$ the  spinor satisfying $\gamma^{\hat r\hat  \theta\hat  \phi} \epsilon^i=\epsilon^i$. From the $\delta \psi_\pm^i$ variations we find that $\epsilon^i$ is independent of $x^\pm$, and that there is another solution of the form\foot{The $\lambda^i$ are the enhanced supersymmetries of the near-horizon region. 
This equation expresses them in terms of the Lie derivative with respect to an $SL(2,{\bf R})$ Killing vector acting on $\epsilon^i$. }   \eqn\sspin{\lambda^i= -{x^+ \over l}\epsilon^i +\sqrt{l\over r}\gamma^{\hat +\hat r}\epsilon^i.} 
Solving the angular variations for $\epsilon^i$ gives \eqn\fulle{\epsilon^i=({r\over l})^{1/4} e^{{\theta \over 2}\gamma^{\hat \phi}} e^{-{\phi \over 2} \gamma^{\hat \theta\hat  \phi}} \epsilon^i_0,} where $\epsilon_0$ is a constant spinor which satisfies $\gamma^{\hat r\hat \theta\hat \phi}\epsilon^i_0 = \epsilon^i_0$. In addition, the  $\lambda^i$ given in terms of $\epsilon^i$ in \sspin\ remain solutions as well since these angular variations commute with $\gamma^{\hat +\hat r}$.  So all in all we have 16 near horizon supersymmetries. 

\subsec{Killing vectors}

In order to determine the complete supergroup, we need to understand the action of the (right-handed) $SL(2,{\bf R})$ bosonic symmetries on the Killing spinors.  The $SL(2,{\bf R})$ Killing vectors  are \eqn\slop{L_{-1}=l\partial_+, \qquad L_0=-x^+ \partial_+ +r \partial_r, \qquad L_1 = {(x^+)^2 \over l} \partial_+ - {2x^+ r \over l}\partial_r +{4l^2 \over r}\partial_-.} Using these we find that\foot{With the action defined via the Lie derivative  ${\cal L}_K\epsilon =K^\mu\nabla_\mu \epsilon +{1 \over 4}\p_\mu K_\nu \gamma^{\mu\nu}\epsilon$. } \eqn\slrep{ L_0 \epsilon^i = {1\over 2}\epsilon^i, \qquad L_0 \lambda^i=-{1\over 2}\lambda^i ,\qquad L_{1} \epsilon^i = \lambda^i, \qquad L_{-1} \lambda^i=-\epsilon^i ,} which identifies $\epsilon^i$ and $\lambda^i$ respectively with the $-\ha$ and $+\ha$ modes of $G$ obeying 
$[L_m,G_r]=({m\over 2} -r)G_{m+r}$.

Similarly the $SU(2)$ action is generated by \eqn\suop{J^3_0=-i \partial_\phi, \qquad J^{\pm}_0 = e^{\pm i \phi} (-i \partial_\theta \pm \cot\theta \partial_\phi ).} Since $\gamma^{\hat r\hat \theta\hat \phi}$ and $\gamma^{\hat \theta\hat  \phi}$ commute we can define \eqn\suproj{\gamma^{\hat \theta\hat \phi}\epsilon^i_{0}=\mp i\epsilon^i_{0},} and it is easy to check that these satisfy \eqn\surep{J^3_0\epsilon^i=\pm{1\over 2} \epsilon^i.} Suppose we start with a constant spinor obeying $\gamma^{\hat \theta\hat \phi}\epsilon_0=-i \epsilon_0$ as well as $\epsilon_0 =-i\gamma^{\hat 0\hat \theta}\epsilon_0^*$, and normalized to $\epsilon_0^\dagger \epsilon_0={1\over 4}$. Then we can define \eqn\spins{\eqalign{\xi^1_-&=({r\over l})^{1/4} e^{{\theta\over 2} \gamma^{\hat \phi}}e^{-{i\over 2}\phi} (\gamma^{\hat \theta} \epsilon_0),\cr 
\xi^1_+&=({r\over l})^{1/4} e^{{\theta\over 2} \gamma^{\hat \phi}}e^{{i\over 2}\phi} \epsilon_0,\cr
\xi^2_-&=({r\over l})^{1/4} e^{{\theta\over 2} \gamma^{\hat \phi}}e^{-{i\over 2}\phi} (-\gamma^{\hat 0} \epsilon_0^*),\cr
\xi^2_+&=({r\over l})^{1/4} e^{{\theta\over 2} \gamma^{\hat \phi}}e^{{i\over 2}\phi} (-\gamma^{\hat 0\hat \theta} \epsilon_0^*), }} where $\xi^a$ is a ${\bf 2}$ of $SU(2)$, $J^\pm_0 \xi^a_{\pm}=0$ and $J^\pm_0 \xi^a_{\mp}=\xi^a_\pm$. We can organize these into symplectic-Majorana Killing spinors \eqn\smks{\eqalign{&\epsilon^{(1)}=\pmatrix{\xi^1_+ \cr -\xi^2_- \cr 0 \cr 0},\quad \epsilon^{(2)}=\pmatrix{\xi^2_+ \cr -\xi^1_- \cr 0 \cr 0},\quad \epsilon^{(3)}=\pmatrix{\xi^1_- \cr \xi^2_+ \cr 0 \cr 0}, \quad \epsilon^{(4)}=\pmatrix{ -\xi_-^2 \cr -\xi^1_+ \cr 0 \cr 0}, \cr &\epsilon^{(5)}=\pmatrix{0\cr 0\cr \xi^1_+ \cr -\xi^2_- },\quad \epsilon^{(6)}=\pmatrix{0\cr 0\cr \xi^2_+ \cr -\xi^1_-},\quad \epsilon^{(7)}=\pmatrix{0\cr 0\cr \xi^1_- \cr \xi^2_+}, \quad \epsilon^{(8)}=\pmatrix{0\cr 0\cr -\xi^2_- \cr - \xi^1_+}, }} where each $\epsilon^{(I)}$ transforms as a ${\bf 4}$ of $Sp(4)$ by left-multiplication (see appendix A). We will identify these with $G^I_{-\ha}$, $I=1,\dots,8$. Following the same procedure we can define \eqn\ets{\eta^a_\pm = -{x^+\over l} \xi^a_\pm + \sqrt{l\over r}\gamma^{\hat +\hat r} \xi^a_\pm} and group them into symplectic-Majorana ${\bf 4}$'s which will be identified with $G^I_\ha$.

\subsec{Supercharge commutators}

Commutators of supercharges can be expressed as fermion bilinears involving the corresponding Killing spinors. In particular, \FujitaKV\ determines \eqn\twog{\eqalign{ \{G^I_r, G^J_s \} \sim &\Omega_{ij} \left( ({\bar \epsilon}^{(I)}_r)^i \gamma^{\mu} (\epsilon^{(J)}_s)^j +({\bar \epsilon}^{(J)}_s)^i \gamma^{\mu} (\epsilon^{(I)}_r)^j \right) \partial_{\mu} \cr &+\left( ({\bar \epsilon}^{(I)}_r)_i \gamma^{\hat \theta\hat \phi} (\epsilon^{(J)}_s)^j +({\bar \epsilon}^{(J)}_s)_i \gamma^{\hat \theta\hat \phi} (\epsilon^{(I)}_r)^j \right),}} where $\epsilon^{(I)}_{-\ha}=\epsilon^{(I)}$ and $\epsilon^{(I)}_\ha=\lambda^{(I)}$. The first line of \twog\ involves the spacetime Killing vectors of $SL(2,{\bf R}) \times SU(2)$ and the second involves the the generators of $Sp(4)$. Using our previous normalizations and the notation in appendix A, we find \eqn\ggl{\{G^I_{\pm \ha},G^J_{\pm \ha}\}=-2\delta^{IJ} L_{\pm 1} } for $I,J=1,\dots,8$. Also,
\eqn\gg{\{G^{I_1}_\ha,G^{J_1}_{-\ha}\}=\pmatrix{-2 L_0 & -2iJ^3_0+i A_3 & 2iJ^2_0+iA_1 & 2iJ^1_0 +iA_2 \cr 2iJ^3_0-iA_3 & -2L_0 & 2iJ^1_0-iA_2 & -2iJ^2_0 +iA_1 \cr -2iJ^2_0 -iA_1 & -2iJ^1_0+iA_2 & -2L_0 & -2iJ^3_0-iA_3 \cr -2iJ^1_0-iA_2 & 2iJ^2_0 -iA_1 & 2iJ^3_0+iA_3 & -2L_0 },} where $I_1,J_1=1,\dots, 4$. If $I_2,J_2=5,\dots,8$, the same table arises with $A_\alpha$ replaced by $C_\alpha$. If $I_1=1,\dots,4$, and $J_2=5,\dots,8$,
\eqn\ggg{\{G^{I_1}_\ha,G^{J_2}_{-\ha}\}=\pmatrix{iB_4 & iB_3 & iB_1 &iB2 \cr -iB_3 & iB_4 & -iB_2 & iB_1 \cr -iB_1 & iB_2 & iB_4 & -iB_3 \cr -iB_2 & -iB_1 & iB_3 &iB_4}.} These are just the commutation relations of $\osp$, written below in a more compact form (assuming we rotate $G\to iG$): \eqn\ospalg{\eqalign{ \{G^I_r,G^J_s\} &=2L_{r+s} \delta^{IJ}+(r-s)(t_\alpha)^{IJ}J^\alpha_0 + (r-s) (\rho_A)^{IJ}R^A_0 \cr \left[ L_m, G^I_s \right]&= ({m\over 2}-s)G^I_{m+s}, \cr \left[ R^A_0,G^I_r\right]&=(\rho^A)^{IJ}G^J_r \cr \left[J^\alpha_0,G^I_r\right]&= (t^\alpha)^{IJ} G^J_r,}} where $t^\alpha$ and $\rho^A$ are the representation matrices for $SU(2)$ and $Sp(4)$ respectively, and $R^A$ are the generators of $Sp(4)$. In the first two lines of \ospalg, it should be understood we have only computed the global part of the superalgebra.

\newsec{Towards an exact worldsheet CFT}
In this section, we review and point out that the old results of \GiddingsWN\ may be
relevant to the problem of finding an exact worldsheet dual.  We note that with the obvious adaptation of the GSO projection used in \GiddingsWN\ one does not realize the needed 16 supercharges, so 
something more is needed to get a fully viable candidate for the worldsheet CFT.

\subsec{4d heterotic black monopoles} Heterotic string theory in
four dimensions contains macroscopic black hole solutions \GarfinkleQJ\ with
magnetic charges lying in a $U(1)$ subgroup of $E_8\times E_8$.
Since the charges are associated with the left-moving sector of the
worldsheet, such solutions are generically non-supersymmetric. The
near horizon region is the product of 2D Minkowski space with  a
linear dilaton and an $S^2$ threaded with magnetic flux.

For every classical solution there should be a corresponding
worldsheet CFT. In this case the CFT is rather subtle but was
eventually found in \GiddingsWN. While $S^3$ factors such as those arising
in the near horizon for the NS5-brane are easily recognized as
$SU(2)$ WZW models (which have $\su_L$ and $SU(2)_R$ current
algebras corresponding to the $S^3$ isometry group) it is harder to
see where an $S^2$ horizon comes from (which has only one $SU(2)$
isometry). It turns out that it is given by an asymmetric orbifold
of level $k=2|Q^2-1|$ WZW model of the
form\eqn\axcv{ { SU(2)_{2|Q^2-1|} \times SU(2)_{2|Q^2-1|} } \over {\bf Z}_{2Q+2},}where $Q$ is the
monopole charge. \axcv\ can be viewed as a two sphere with a left
and a right fiber $U(1)_L$ and $U(1)_R$. The $U(1)_L$ fiber comes
from the $U(1)$ subgroup of $E_8\times E_8$ and the Chern class of
the fibration is determined by the monopole charge. On the right,
one has two fermions which are superpartners of the the two
coordinates of the $S^2$ horizon and live in the tangent bundle.
These can be bosonized to a $U(1)_R$ boson which also has a
nontrivial fibration. The total space of the $S^2$ horizon together
with its bosonized right-moving superpartners and the left-moving
current $U(1)_L$ boson was shown in \GiddingsWN\ to be given by \axcv, with
a specified action for the ${\bf Z}_{2Q+2}$ quotient. \subsec{5d
monopole-heterotic strings} We wish to consider two modifications of
the construction of \GiddingsWN. First, by trading a compact dimension for
a trivial flat dimension, we can uplift the CFT to one describing a
monopole string in five dimensions. Second, we replace the 3d
$M^2\times(linear~~dilaton)$ factor with a $(0,1)$ $\rl_{k+4}$ WZW
model\foot{The supersymmetric right side contains bosonic level
$k+4$ $\rl$ current $j^A$ and a supersymmetric level $k+2$ $\rl$
current $J^A$. } (representing an $AdS_3$ factor) with the same
central charges $(c_L,c_R)=(3+{6 \over k+2},{9\over 2}+{6 \over k+2})$
and constant dilaton.

   The presence of $H$ flux on the $AdS_3$ factor indicates that
the monopole string also carries fundamental string charge. The
number $N$ of heterotic strings behaves as \eqn\dxa{N\sim \int_{S^2\times M_5}e^{-2\Phi}*H\sim {k \over g_5^2}.} We wish to
have weakly coupled string theory so $N$ must be large.

\subsec{Q=0: the heterotic string near-horizon}
   An intriguing feature of the construction of \GiddingsWN\ is that it is
   nonsingular for the case $Q=0$ which corresponds to $k=2$.
   This case was referred to as the ``neutral remnant" in \GiddingsWN. $k=2$
   can be described by 3 left and 3 right free fermions, and
   the ${\bf Z}_2$ quotient in \axcv\ acts purely on the left as a $2\pi$
   rotation. One then expects that with the modifications of the previous
     subsection, the case $k=2$ corresponds to the near-horizon geometry of
     $N$ strings.
However, to define the theory we must
specify the GSO projection (with the $SL(2,{\bf R})$ factor there seems to
be more than one way to do this), and the obvious adaptation of the one given in \GiddingsWN\ does not give the needed spacetime supersymetries. Possibly a different value of $k$\foot{The construction of \GiveonNS\ naively indicates that 
the value $k=0$ (which gives bosonic $SL(2,R)$ currents at level 4), gives the desired spacetime central charge, but this construction requires modifications for a nonlinear  superconformal algebra. } or modified GSO will give the desired theory. 

\centerline{\bf Acknowledgements}
   This work was supported in part by DOE grant DE-FG02-91ER4064.  AS would like to thank Atish Dabholkar and Sandip Trivedi for collaboration at an earlier stage on this problem. 
We are grateful to J. Harvey, P. Kraus,  D. Kutasov, F. Larsen, J. Maldacena, G. Moore  and Xi Yin for discussions and correspondence. 

\appendix{A}{$Sp(4)$}

An element $g\in Sp(4)$ satisfies $g^\dagger g=1$ and $g^T \Omega g=\Omega$. If we parameterize $g$ as $e^{iM}$, $M \in {\bf sp}(4)$, then $M=M^\dagger$ and $M^T\Omega+\Omega M=0$. This determines \eqn\spn{{\bf sp}(4)={\rm span}_{\bf R}\{ A_1,A_2,A_3,B_1,B_2,B_3,B_4,C_1,C_2,C_3\}.} Writing these in $2\times 2$ blocks, \eqn\abc{\eqalign{&A_\alpha=\pmatrix{\sigma^\alpha & 0 \cr 0 &0}, \quad C_\alpha=\pmatrix{0&0\cr 0&\sigma^\alpha}, \cr &B_\alpha=\ha \pmatrix{0& i^{\delta_{\alpha,2}} \sigma^\alpha \cr (i^{\delta_{\alpha,2}} \sigma^\alpha)^\dagger &0},\quad B_4=\ha \pmatrix{0 & i \cr -i & 0 }, }} where $\sigma^\alpha$ are the Pauli matrices
\eqn\sig{
\sigma^1 = \pmatrix{ 0 & 1 \cr 1 & 0 }, \quad  \sigma^2 = \pmatrix{ 0 & -i \cr i & 0 },  \quad  \sigma^3 = \pmatrix{ 1 & 0 \cr 0 & -1 } .}

\listrefs

\end